\documentclass[prl,twocolumn,english,showpacs]{revtex4-1}
\usepackage{graphicx}
\usepackage{amssymb}
\usepackage[rightcaption]{sidecap}

\begin{document}

\title{Scale invariance and viscosity of a two-dimensional Fermi gas}

\author{Enrico Vogt, Michael Feld, Bernd Fr{\"o}hlich, Daniel Pertot}
\author{Marco Koschorreck}
\email[E-mail: ]{mk673@cam.ac.uk}
\author{Michael K{\"o}hl}

\affiliation{Cavendish\,Laboratory,\,University~of Cambridge, JJ Thomson Avenue, Cambridge CB3 0HE, United Kingdom}

\begin{abstract}
We investigate collective excitations of a harmonically trapped two-dimensional Fermi gas from the collisionless (zero sound) to the hydrodynamic (first sound) regime. The breathing mode, which is sensitive to the equation of state, is observed with an undamped amplitude at a frequency two times the dipole mode frequency for a large range of interaction strengths and different temperatures. This provides evidence for a dynamical SO$(2,1)$ scaling symmetry of the two-dimensional Fermi gas. Moreover, we investigate the quadrupole mode to measure the shear viscosity of the two-dimensional gas and study its temperature dependence.
\end{abstract}


\date{\today}

\maketitle

Scale invariant behaviour plays an important role in many branches of physics. It is encountered both in fluctuations near critical points of phase transitions \cite{ZinnJustin2002} and in particle physics when masses become unimportant at high energies \cite{Coleman1985}. In non-relativistic quantum mechanics, scale invariance implies that the Hamiltonian $H(x)$ scales under a dilatation of the spatial coordinate $x\rightarrow \lambda x$ according to $H(\lambda x) \rightarrow H(x)/\lambda^2$. This scaling symmetry allows one to make fundamental statements about thermodynamic properties. For example, even in a strongly interacting system in $D$ dimensions, pressure $P$ and energy density $\varepsilon$ are related by the same simple equation of state $P = 2\varepsilon/D$ as an ideal gas. One such system, which is of great interest in the ultracold atom, heavy-ion, and nuclear astrophysics communities, is the two-component Fermi gas in three dimensions interacting via a zero-range potential with a scattering length $a_{3D}$. At unitarity ($a_{3D}\rightarrow \infty$) it is scale- and conformally-invariant with universal properties \cite{Ohara2002,Ho2004b,Son2007,Nishida2007,Enss2011}. For example, the equation of state at zero temperature becomes $\mu \propto E_F$, in which $\mu$ is the chemical potential and $E_F$ is the Fermi energy, and the bulk viscosity $\zeta$ vanishes for arbitrary temperatures.

In two-dimensional systems scaling behaviour is more subtle. When the two-dimensional scattering length $a_{2D}$ \footnote{Following D. Petrov and G. Shlyapnikov, Phys. Rev. A 64, 012706 (2001), we use the definition $E_B=\hbar^2/ma_{2D}^2$ in which $E_B$ is the binding energy of the dimer state.} approaches infinity, the gas becomes non-interacting. This implies that at zero temperature one finds $\mu=E_F\propto n_{2D}$ and hence the gas is trivially scale invariant. Here, $n_{2D}$ is the density. At finite interaction strength, i.e., finite values of $a_{2D}$, the two-body scattering amplitude in two dimensions $f(q)=\frac{4 \pi}{-\ln(q^2a_{2D}^2)+i\pi}$ is momentum dependent. Evaluating $f(q)$ at a characteristic momentum, for example the Fermi wave vector $k_F$, leads to a density dependent coupling strength \cite{Bloom1975}. In a quantum field theoretical model of the superfluid Bose gas, it has been pointed out that this could give rise to a quantum anomaly that breaks scale invariance \cite{Olshanii2010}.

In the presence of an isotropic harmonic trap, the scale invariance of the homogeneous system is replaced by a dynamical SO$(2,1)$ scaling symmetry \cite{Pitaevskii1997}. The SO$(2,1)$ group, or ``Lorentz'' group, is the group of rotations in 2+1 dimensional space-time. For the trapped gas, the SO$(2,1)$ symmetry results in an excitation spectrum with modes spaced by exactly $2 \omega_\perp$\cite{Pitaevskii1997,Werner2006}. Here, $\omega_\perp$ denotes the trap frequency of the weakly confined axes. This generates a hydrodynamic breathing mode at a frequency $\omega_B=2\omega_\perp$, independent of the interaction strength. Moreover, the mode frequency is independent of amplitude and the breathing mode should be undamped. However, the quantum anomaly resulting from the density-dependent coupling strength has been predicted to shift the hydrodynamic breathing mode of a Bose gas by $\delta\omega_B/\omega_B = \frac{1}{4\sqrt{\pi}} a_{3D}/l_z$ for $a_{3D}/l_z\ll 1$\cite{Olshanii2010} where $l_z$ denotes the extension of the gas in the strongly confined direction.

\begin{SCfigure*}
\includegraphics[width=.7\textwidth,clip=true]{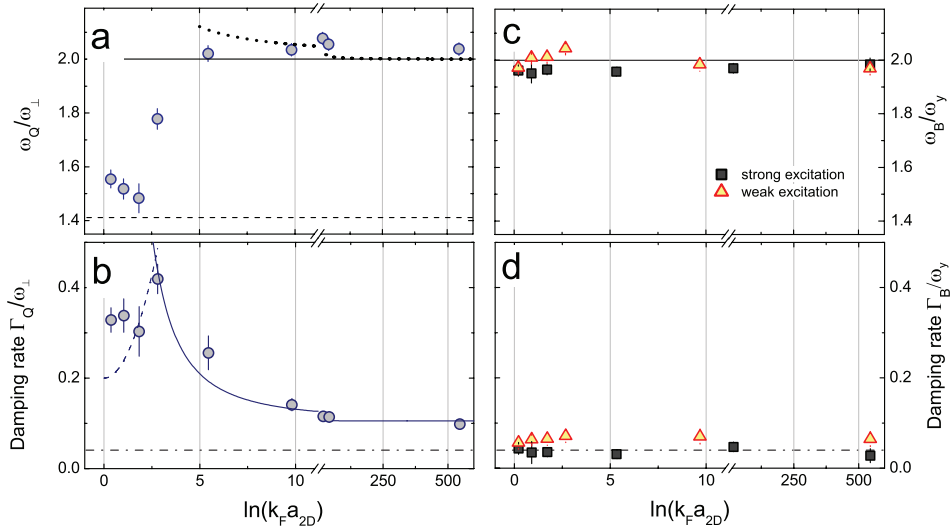}
\caption{(Color online) \textbf{a:} Frequency of the quadrupole mode at $T/T_F=0.47$ with $E_F=h\times(8.2\pm0.7)\,$kHz. The dotted line shows the theoretical prediction of \cite{Ghosh2002} in the collisionless limit and the dashed line at $\omega_Q/\omega_\perp=\sqrt{2}$ is the hydrodynamic case. \textbf{b:} Damping of the quadrupole mode. The solid line shows the fit for  $\Gamma_0$ (zero sound) and the dashed line for $\Gamma_1$ (first sound) - see text. The dash-dotted line is the damping rate of the dipole mode. \textbf{c:} Frequency of the breathing mode. For the strong excitation we have $T/T_F=0.37$ and $E_F=h\times(5.4\pm0.8)$\,kHz and for the weak excitation $T/T_F=0.42$ and $E_F=h\times(8.1\pm 1.1)$\,kHz. \textbf{d:} Damping of the breathing mode. The dash-dotted line is the damping rate of the dipole mode.}
\label{fig1}
\end{SCfigure*}

In this manuscript, we study an interacting two-dimensional Fermi gas using collective modes to investigate scale invariance and viscosity. We tune the ratio $a_{3D}/l_z$ between $-3$ and $0$, which provides us with access to the hydrodynamic and the collisionless regimes. Previous experiments using ultracold atomic gases to study scale invariance of a two-dimensional Hamiltonian were limited to weakly interacting bosons in a regime where $0<a_{3D}/l_z\ll 1$ \cite{Hung2011,Yefsah2011} and to weakly interacting three-dimensional gases with a highly elongated symmetry \cite{Chevy2002}.

In our experiment \cite{Frohlich2011,Feld2011}, we create two-dimensional Fermi gases of $^{40}$K atoms in a 50/50 mixture of the $|F=9/2,m_F=-9/2\rangle$ and $|F=9/2,m_F=-7/2\rangle$ states of the hyperfine ground state manifold. The Fermi gas is loaded into the standing wave potential of an optical lattice to create an array of two-dimensional gases. The trapping frequency along the strongly confined direction is $\omega_z=2 \pi \times 78$\,kHz and the radial trapping frequency is $\omega_\perp=\sqrt{\omega_x \omega_y}\simeq 2\pi\times 125$\,Hz with an anisotropy $\epsilon=|\omega_x-\omega_y|/2\omega_\perp$ below $2\%$. Along the axial direction we populate approximately 30 layers of the optical lattice potential with an inhomogeneous peak density distribution of typically $2 \times 10^3$ atoms per spin state per 2D gas at the center. We tune the interactions by applying a magnetic field close to the Feshbach resonance at 202.15\,G \footnote{We use the Feshbach resonance parameters from N. Strohmaier {\it et al.}, Phys. Rev. Lett. {\bf 104}, 080401 (2010).}.

First, we study the quadrupole mode of the two-dimensional Fermi gas in order to identify the collisionless and the hydrodynamic regimes. The quadrupole mode has the incompressible velocity field $\textbf{v}_Q(\textbf{r})=b[x\hat{\textbf{e}}_x-y\hat{\textbf{e}}_y]\cos(\omega_Q t)$ with a constant $b$, and corresponds to a surface mode of the gas. In the collisionless limit, surface modes are analogous to zero sound. At zero temperature the quadrupole mode frequency is predicted to be $\omega_Q=\sqrt{2(2-\tilde{g})/(1-\tilde{g})} \omega_\perp$ with $\tilde{g}=1/\ln(k_Fa_{2D})$ \cite{Ghosh2002}. A collective hydrodynamic mode, in contrast, corresponds to a first sound mode. The frequency of the hydrodynamic quadrupole mode $\omega_Q=\sqrt{2}\omega_\perp$ \cite{Wen2007,Klimin2011} is independent of the equation of state because its incompressible flow pattern prevents a change of the internal energy during the oscillation \cite{Bulgac2005}.

We excite the quadrupole mode by adiabatically introducing a small anisotropy to the two-dimensional harmonic oscillator potential using additional laser beams while maintaining $\omega_x\times \omega_y\approx const.$ and then abruptly returning to the original trapping configuration. The atomic cloud oscillates freely in this potential for up to 14\,ms until we switch it off and take an absorption image after 12\,ms of time of flight. The velocity amplitude of the excitation is $10\%$ of the Fermi velocity $v_F =\hbar k_F /m$. We determine the radius of the cloud in the x- and the y-direction and fit their difference to measure $\omega_Q$ \cite{Altmeyer2007}. Owing to the change in gravitational sag during the excitation, we simultaneously excite small dipole (center-of-mass) oscillations primarily in the vertical y-direction. We use these dipole oscillations to calibrate $\omega_{x,y}$, of which $\omega_y$ has the smaller error because of the larger oscillation amplitude. The decay rate of the bare dipole mode $\Gamma_D/\omega_\perp=0.04\pm0.01$ is most probably caused by the weak anharmonicity of our Gaussian trapping potential.

In Figure \ref{fig1}a we show the quadrupole mode frequency $\omega_Q$ of the two-dimensional Fermi gas. For large values of the interaction parameter $\ln(k_Fa_{2D})$ we are in the collisionless regime and observe $\omega_Q\simeq 2\omega_\perp$, in agreement with the theoretical expectation \cite{Ghosh2002}. As we increase the interaction strength, i.e., decrease the value of $\ln(k_Fa_{2D})$, we enter the hydrodynamic regime. This transition is marked by a sharp decrease of the collective mode frequency to $\omega_Q \simeq\sqrt{2}\omega_\perp$. Theoretically, we expect the transition from the collisionless to the hydrodynamic regime when the collision rate $\gamma_0$ equals the mode frequency $\omega_Q$. We estimate $\gamma_0=-\hbar n_{2D} \textrm{Im}[f(k_F)]/m$ using the optical theorem for the scattering cross section $\sigma=-\textrm{Im}[f(q)]/q$. For very small deviations from the equilibrium distribution, the collision rate $\gamma_0$ is suppressed by a factor $(T/T_F)^2<1$ owing to the restriction of phase space because of Pauli's exclusion principle. Using our average density $n_{2D} \approx 6\times 10^{12}$m$^{-2}$ and temperature $T/T_F=0.47$ we estimate the transition from the hydrodynamic to the collisionless regime at an interaction parameter $\ln(k_Fa_{2D})\approx 3$, in good agreement with the observed mode frequency change.

In Figure \ref{fig1}b we show the damping rate $\Gamma_Q$ of the quadrupole mode. The zero sound mode in the collisionless regime is damped by collisions which disrupt the coherent quasiparticle motion. Hence, the mode damping rate $\Gamma_0$ scales proportional to the normalized collision rate $\gamma_0/\omega_Q$ with an asymptotic behaviour $\Gamma_0 \propto \left[\ln(k_Fa_{2D})\right]^{-2}$, shown as the solid line. The damping rate reaches a constant offset of $0.1\,\omega_\perp$, even for the non-interacting gas, which we will investigate in more detail below. For the first sound mode in the hydrodynamic regime the situation is opposite: collisions are necessary to establish local equilibrium and first sound is damped by the deviation from this, hence $\Gamma_1\propto \omega_Q/\gamma_0$, which asymptotically is $\Gamma_1 \propto 1+\left[2\ln(k_Fa_{2D})/\pi\right]^{2}$ (dashed line). In the fit of the proportionality constant we have used the same constant offset as for the non-interacting gas. In between the two extremes, the damping rate peaks at the transition from the collisionless to the hydrodynamic regime.

Having identified the hydrodynamic and collisionless regimes, we now turn our attention to the breathing mode and the question of scale invariance of the two-dimensional Fermi gas. The velocity field of the breathing mode is $\textbf{v}_B(\textbf{r})=b[x\hat{\textbf{e}}_x+y\hat{\textbf{e}}_y]\cos(\omega_B t)$ with a constant $b$. The breathing mode is excited by adiabatically decreasing the strength of the two-dimensional confinement and then abruptly returning to the original trapping configuration.
After an oscillation time of up to 20\,ms we switch off the confinement and take an absorption image after 12ms time of flight. We identify the breathing mode frequency (Figure \ref{fig1}c) by studying the cloud radii in x- and y-direction as a function of hold time and define $\omega_B=\sqrt{\omega_{B,x}\omega_{B,y}}$. Again, we use the simultaneously excited dipole mode for continuous referencing. We study two sets of data of the breathing mode as a function of interaction strength which differ by a factor of 2 in excitation strength \footnote{For the weak excitation of the breathing mode the spin imbalance was $10\%$ of the total atom number and the data point at $\ln(k_Fa_{2D})=2.7$ has $\epsilon=4.1\%$.}. The weak excitation strength corresponds to $12\%$ modulation of the width after time of flight. We observe that the mode frequency is approximately constant for all interaction strengths and averages at $\omega_B/\omega_y=2.00\pm0.03$ for the weak excitation and at $\omega_B/\omega_y=1.96\pm0.01$ for the strong excitation. We also have not observed any change of the mode frequency with the temperature of the gas in the range $0.37<T/T_F<0.9$.

The independence of the breathing mode frequency from interaction strength, oscillation amplitude, and temperature and the fact that the mode frequency is very close to $\omega_B=2\omega_\perp$ suggest that the gas exhibits a dynamical SO$(2,1)$ scaling symmetry. In the hydrodynamic normal regime, our result implies that the equation of state is polytropic and $\mu \propto n_{2D}$. Interestingly, this result is compatible with BCS mean field theory at finite temperature \cite{Loktev2001,Klimin2011}. A more precise determination of the equation of state, which also takes into account logarithmic interaction energy shifts and other beyond mean-field effects, could be obtained from Quantum-Monte Carlo calculations, which so far are only available for superfluids at zero temperature \cite{Bertaina2011}.

The damping of the breathing mode (Figure \ref{fig1}d) differs fundamentally from the quadrupole mode, because it is very small and independent of the interaction strength. In order to understand the undamped behaviour of the breathing mode, we employ a simplified model based on classical hydrodynamics. Energy dissipation of the velocity field $\textbf{v}_B(\textbf{r})$, and hence damping of the coherent motion of particles, is caused by the viscosity of the gas. The energy dissipation rate $\dot{E}$ follows from the two-dimensional stress tensor \cite{Landau1976} $\dot{E} =-\frac{1}{2}\int d^2r\, \eta(\textbf{r})(\partial_kv_i +\partial_iv_k-\delta_{ik} \nabla\cdot\textbf{v})^2-\int d^2r\, \zeta(\textbf{r})(\nabla\cdot \textbf{v})^2$ with the shear viscosity $\eta(\textbf{r})$ and the bulk viscosity $\zeta(\textbf{r})$. The time averaged energy dissipation rate is $\langle \dot{E}\rangle_t=-2b^2\int d^2r\, \zeta(\textbf{r})$, entirely determined by the bulk viscosity. The amplitude damping of the mode is $\Gamma= \langle \dot{E}\rangle_t/2 \langle E \rangle_t$, using the time-averaged mechanical energy $\langle E \rangle_t=\frac{mb^2}{2}\int d^2r\, r^2 n(\textbf{r})$. In our data, the damping rate of the breathing mode averages near $\Gamma_B/\omega_y \simeq 0.05$, equal to the damping of the dipole mode which is dominated by technical limitations (e.g. dephasing due to trap anharmonicities) rather than by viscous forces. Additionally, we observe the absence of damping for different excitation amplitudes and temperatures.

\begin{figure}
 \includegraphics[width=0.8\columnwidth,clip=true]{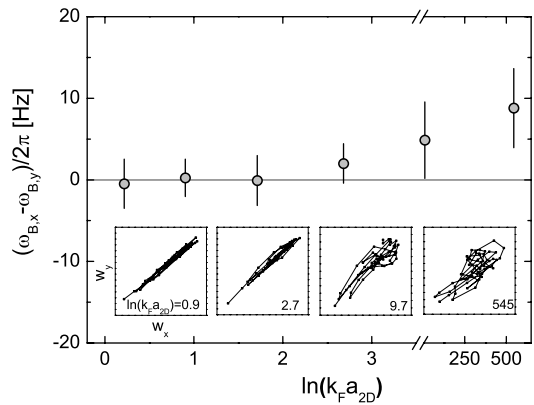}
 \caption{Difference between the breathing mode frequencies of the two different axes at the collisionless-hydrodynamic crossover in a slightly anisotropic trap. The insets show the correlation plot of the cloud widths along the two axes.}
 \label{fig4}
\end{figure}

As a side remark, we observe the transition from the collisionless to the hydrodynamic regime also directly for the breathing mode. In a slightly anisotropic trap ($\epsilon<0.02$) in the collisionless regime, the oscillation frequencies of the widths of the cloud are split by $2\epsilon \omega_B$, corresponding to two independent modes (see Figure \ref{fig4}). When the interaction strength is increased into the hydrodynamic regime the two mode frequencies lock together and the whole cloud undergoes a collective breathing mode at a single frequency. The occurrence of this ``mode-locking'' coincides with the values of $\ln(k_Fa_{2D})$, where we observe the collisionless-hydrodynamic transition of the quadrupole mode. As insets we show correlation plots between the widths of the cloud demonstrating the change from nearly uncorrelated into highly correlated motion.

\begin{figure}
 \includegraphics[width=\columnwidth,clip=true]{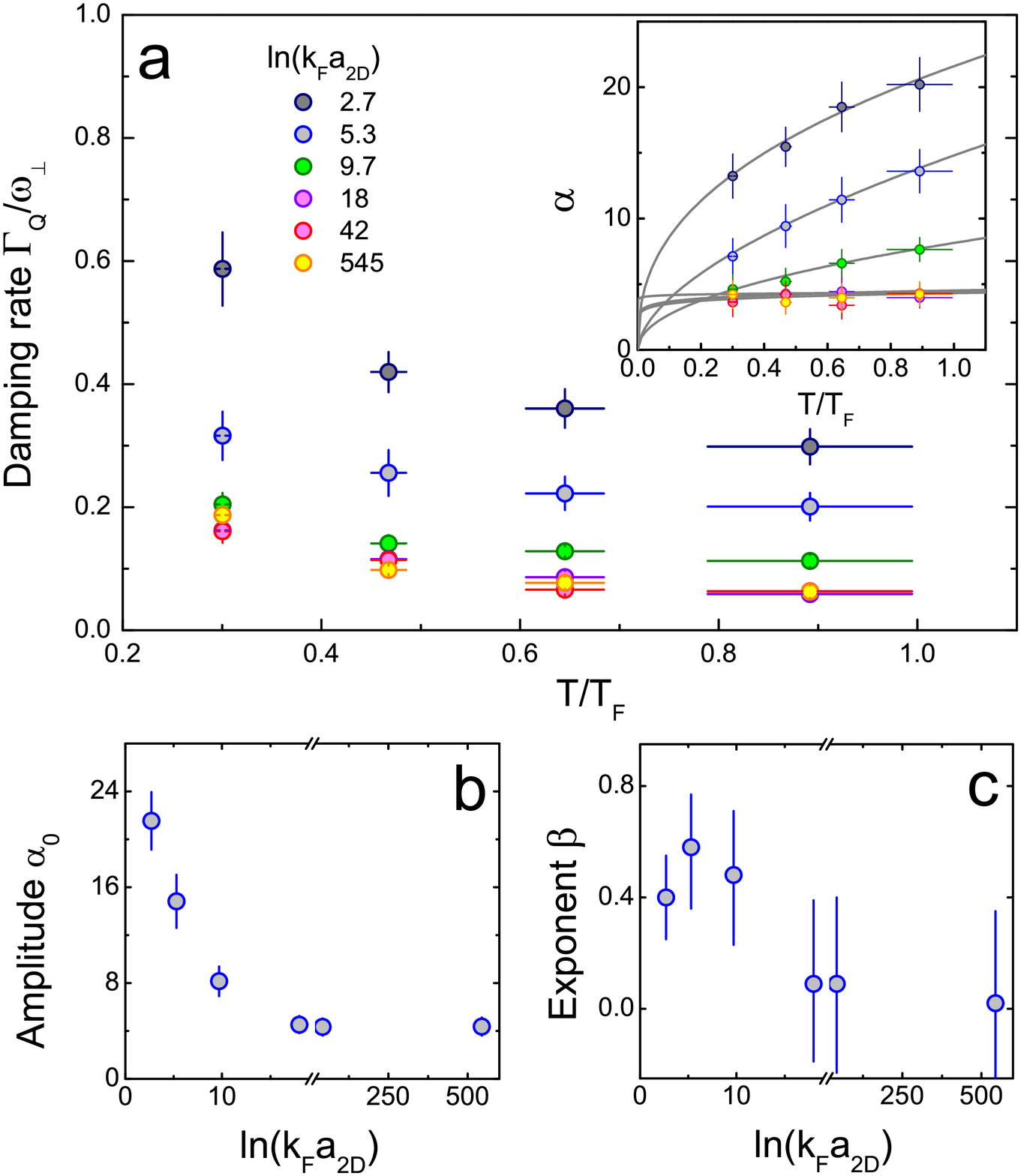}
 \caption{(Color online) Temperature dependent damping of the quadrupole mode. \textbf{a:} Damping rate as a function of temperature for various interaction strengths. $E_F/h$ for the data sets are (from lowest to highest temperature): 6.4\,kHz, 8.2\,kHz, 9.0\,kHz, and 9.1\,kHz. Inset: Derived values of $\alpha(T/T_F)$ and power-law fits.  \textbf{b:} Viscosity amplitude $\alpha_0$. \textbf{c:} Power law exponent $\beta$.}
 \label{figvisc}
\end{figure}

We now turn our attention to the temperature dependent damping of the quadrupole mode. The viscosity of a strongly interacting gas plays an important role in the investigation of the ratio of viscosity to entropy density predicted by the anti-de Sitter/Conformal Field Theory correspondence, which so far has been primarily investigated in three dimensions \cite{Kovtun2005,Cao2011,Cao2011b}. In two dimensions, the shear viscosity of the Fermi gas and its temperature dependence have been theoretically investigated only in specific limits \cite{Novikov2006,Hofmann2011}, none of which correspond to our experimental situation. Our data for the temperature dependence of the damping rate $\Gamma_Q$ are displayed in Figure \ref{figvisc}a for various interaction strengths. We use the same hydrodynamical model as above in order to link the shear viscosity to the damping rate, which neglects possible temperature gradients as well as quantum statistics. The time averaged dissipation rate of the quadrupole mode is $\langle \dot{E}\rangle_t=-2b^2\int d^2r\, \eta(\textbf{r})$. Using the parametrization $\eta(\textbf{r})=\hbar n_{2D}(\textbf{r}) \alpha(T/T_F)$ with a dimensionless viscosity $\alpha(T/T_F)$ \cite{Bruun2007}, we obtain $\alpha(T/T_F)=  m\langle r^2\rangle \Gamma_Q/2 \hbar$. We determine the rms-radius of the cloud $\sqrt{\langle r^2\rangle}$ numerically for the non-interacting gas. We fit our data with a power law $\alpha(T/T_F)=\alpha_0\times (T/T_F)^\beta$, which is known to provide the high-temperature scaling in three dimensions. The extracted amplitudes $\alpha_0$ and exponents $\beta$ are shown in Figures \ref{figvisc}b and \ref{figvisc}c, respectively. For $\ln(k_Fa_{2D})>10$, i.e. deep in the collisionless regime, we observe no significant temperature dependence (i.e. $\beta\simeq 0$) and hence a constant $\alpha$, which at least in part could be due to technical limitations such as anharmonicities. In the weakly interacting regime, the temperature dependence of the viscosity is significant with $\beta \simeq 1/2$. In contrast to three dimensions, where the viscosity has been investigated intensively \cite{Kovtun2005,Cao2011,Cao2011b,Enss2011}, no theoretical prediction exists yet for this parameter range of a two-dimensional Fermi gas.

In conclusion, we have studied collective oscillations of a two-dimensional trapped Fermi gas in the collisionless and the hydrodynamic regimes. We have observed the existence of a breathing mode at two times the trap frequency, which is invariant against interaction strength, amplitude of the excitation, and temperature. Moreover, this breathing mode is undamped as compared to the dipole mode. These observations suggest a dynamic SO$(2,1)$ scaling symmetry of the trapped two-dimensional Fermi gas. In our parameter range we do not observe indications for a quantum anomaly breaking scale invariance. Using the quadrupole mode, we have additionally studied the temperature-dependence of the shear viscosity of the two-dimensional Fermi gas.

We thank G. Bruun, Y. Castin, M. Olshanii, and W. Zwerger for discussions. The work has been supported by {EPSRC} (EP/G029547/1), Daimler-Benz Foundation (B.F.), Studienstiftung, and DAAD (M.F.).

\end{document}